\documentclass[aps,prb,twocolumn,superscriptaddress,floatfix,10pt,nofootinbib]{revtex4-2}
\usepackage[utf8]{inputenc}

\usepackage{graphicx}
\usepackage{latexsym}
\usepackage{amssymb}
\usepackage{amsmath}
\usepackage{amsfonts}
\usepackage{amsthm}
\usepackage{bm}
\usepackage{caption}
\usepackage{enumitem}
\usepackage[draft=false]{hyperref}
\usepackage{lipsum}
\usepackage{mathtools}
\usepackage{braket}
\usepackage{tikz}
\usetikzlibrary{arrows,decorations.pathreplacing,decorations.markings,arrows.meta,patterns,3d,shapes.geometric}
\usepackage{tikz-cd}
\usepackage{comment}
\usepackage{ragged2e}
\usepackage{multirow}
\usetikzlibrary{arrows}
\usepackage{subcaption}
\usepackage[justification=justified, format=plain, font=small]{caption}
\usepackage[normalem]{ulem} 
\usepackage{makecell}
\usepackage{pifont}

\newcommand{\sfigref}[2]{Fig.\,\hyperref[#1]{\ref{#1}(#2)}}

\def\({\left(}
\def\){\right)}
\def\[{\left[}
\def\]{\right]}
\newcommand{\ie}{\begin{equation}\begin{aligned}}
\newcommand{\fe}{\end{aligned}\end{equation}}

\theoremstyle{definition}
\newtheorem{definition}{Definition}

\newtheorem{lemma}[definition]{Lemma}
\newtheorem{proposition}[definition]{Proposition}
\newtheorem{corollary}[definition]{Corollary}

\usepackage{setspace}
\usepackage{etoolbox}
\makeatletter
\patchcmd{\@footnotetext}
  {\setspace@singlespace}{1}
  {}{}
\makeatother

\hypersetup{
  colorlinks = true,
  urlcolor = blue,
  pdfauthor = {},
  pdftitle = {}
}

\begin{document}
\hfill MIT-CTP/5564

\title{Ground State Degeneracy of Infinite-Component Chern-Simons-Maxwell Theories: Foliated vs.~Non-foliated Fracton Orders}
\date{\today}

\author{Xie Chen}
\affiliation{Department of Physics and Institute for Quantum Information and Matter, \mbox{California Institute of Technology, Pasadena, California 91125, USA}}

\author{Ho Tat Lam}
\affiliation{Center for Theoretical Physics, \mbox{Massachusetts Institute of Technology,  Cambridge, MA 02139 USA}}

\author{Xiuqi Ma}
\affiliation{Department of Physics and Institute for Quantum Information and Matter, \mbox{California Institute of Technology, Pasadena, California 91125, USA}}

\begin{abstract} 
Infinite-component Chern-Simons-Maxwell theories with a block-Toeplitz $K$ matrix provide a vast landscape of gapped and gapless, foliated and non-foliated fracton orders. In this paper, we investigate the ground state degeneracy (GSD) of these theories, classifying distinct behaviors of the GSD as a function of the linear system size, i.e.~the size of the $K$ matrix. We find that the GSD can exhibit exponential or polynomial growth, cyclic variations across a finite set of values, or erratic fluctuations within an exponential envelope. We relate these different patterns to the roots of the determinant polynomial -- a Laurent polynomial associated with the block-Toeplitz $K$ matrix. These roots also play a crucial role in determining whether the theory is gapped or gapless. In addition, we propose a necessary condition for a gapped infinite-component Chern-Simons-Maxwell theory to be a foliated fracton order, based on the porperties of the determinant polynomial.
\end{abstract}

\maketitle

\tableofcontents


\section{Introduction}

Fracton orders \cite{Chamon:2004lew,Haah:2011drr,Vijay:2015mka,Vijay:2016phm} are exotic phases of matter characterized by particles with restricted mobility.  For example, fractons are completely immobile, lineons can move only along a one-dimensional line, and planons can move only within a two-dimensional plane. Another perculiar feature of fracton orders is that their ground state degeneracy (GSD)  depends not only on the topology of the underlying manifold but also on the size and the geometry of the underlining lattice \cite{Shirley:2017suz}.  
The GSD of different fracton orders can have drastically different behaviors. For example, it can grow exponentially or polynomially with the linear system size or fluctuates erratically as the system size varies. See the examples in Table \ref{tab:GSD}.
This sensitive dependence of the GSD on the lattice details is a manifestation of the peculiar UV/IR mixing which is the reason why fracton orders defy a conventional continuum limit \cite{Seiberg:2020bhn,Gorantla:2021bda}. Many of these unconventional properties are closely related to the exotic symmetries associated to the fracton orders \cite{Seiberg:2019vrp,Seiberg:2020bhn,Seiberg:2020wsg,Seiberg:2020cxy,Gorantla:2022eem}. See \cite{Nandkishore:2018sel,Pretko:2020cko} for reviews on fracton orders.

\renewcommand{\arraystretch}{1.4} 
\begin{table*}    
\centering
    \begin{tabular}{|c|c|c|c|c|c|}
         \hline
         Fracton Model & GSD & Pattern of GSD
         \\
         \hline
         X-cube model \cite{Vijay:2016phm} & $2^{6N-3}$ \cite{Vijay:2016phm} & Grow exponentially
         \\
         \hline
         Ising cage-net model \cite{Prem:2018jsn} & $\dfrac{1}{8}\left(9^{3N}+3\times 9^{2N}+15\times 9^N+45\right)$ \cite{Ma_2023} & Grow as a sum of exponentials
         \\
         \hline
        Haah code \cite{Haah:2011drr} &  $2^{r+1}(1+6q_7+6q_{21}+30q_{31}+60q_{63}+\cdots)$ \cite{Haah:2011drr} & Fluctuate erratically
         \\
         \hline
         \makecell{ 2d $U(1)$ vector charge\vspace{1pt}
         \\
         tensor
         gauge theory \cite{Xu_2006,Pretko:2016kxt,Pretko:2016lgv} }  & {$N^2$  \cite{Gorantla:2022ssr}} & Grow polynomially
         \\
         \hline 
         \noalign{\vspace{\normalbaselineskip}}
         \hline
         Tridiagonal iCS Theory & GSD & Pattern of GSD
         \\
         \hline
         $(2,0)$ iCS theory &$2^N$ & Grow exponentially
         \\
         \hline
         $(3,1)$ iCS theory & $\left(\dfrac{3-\sqrt{5}}{2}\right)^N+\left(\dfrac{3+\sqrt{5}}{2}\right)^N-2(-1)^N$ &   Grow as a sum of exponentials
         \\
         \hline
         $(1,2)$ iCS theory & $4\times2^N\sin^2\left(\dfrac{\arccos(-1/4) N}{2}\right)$ & Fluctuate erratically
         \\
         \hline
        $(2,1)$ iCS theory  & $Nq_2+(1-q_2)$ &
        \makecell{Grow polynomially (even $N$)\vspace{1pt}
        \\
        Contant (odd $N$)}
         \\
         \hline
         $(-1,1)$ iCS theory & $q_6+4\sin^2(\pi N/6) (1-q_6)$ & Cycles between finite values
         \\
         \hline
         \end{tabular}
         \captionsetup{justification=Justified}
        \caption{ The GSD of various fracton models on square/cubic lattice and various tridiagonal iCS theories. The iCS theories are labeled by their $(M_0,M_1)$ as defined in \eqref{eq:Kmatrix_pic}. In both cases, we impose periodic boundary conditions. In the former cases, $N$ is the linear size of the lattice while in the latter ones, $N$ is the number of layers. The formula for the Haah code is valid only for $2\leq N\leq 100$. Here, $r = r(N)$ is the largest integer such that $2^r$ divides $N$, and $q_n = q_n(N)$ equals to 1 if $n$ divides $N$ and equals to 0 otherwise. Another fracton model with erratic pattern of GSD was discussed recently in \cite{Gorantla:2022pii}. }
    \label{tab:GSD}
\end{table*}

A novel class of three-dimensional fracton orders described by an infinite-component Chern-Simons-Maxwell (iCS) theory was proposed in \cite{Ma:2020svo}. The theory is made of an infinite number of stacked two-dimensional layers where each layer supports an equal number of two-dimensional $U(1)$ gauge fields $\mathcal{A}_{\mu=0,1,2}^{I,a}$. Here, the index $I$ serves as a discrete coordinate of the transverse $x_3$ direction while the other index $a=1,...,L$ labels different gauge fields on the same layer. The gauge fields have components only in the $x_0,x_1,x_2$ directions. Therefore, charged particles (if localized) have currents only along these three directions, and are hence planons. 
The gauge fields are coupled to each other through an integer-valued symmetric $K$ matrix that gives rise to a Euclidean Lagrangian
\ie\label{eq:action}
\mathcal{L}=\frac{1}{4g^2}\mathcal{F}^{I,a}_{\mu\nu}\mathcal{F}^{\mu\nu}_{I,a}+\frac{i}{4\pi} K_{IL+a,JL+b}\epsilon^{\mu\nu\rho} \mathcal{A}^{I,a}_\mu \partial_\nu \mathcal{A}^{J,b}_\rho~,
\fe
where $\mathcal{F}^{I,a}_{\mu\nu}$ is the field strength of $\mathcal{A}^{I,a}_\mu$. Imposing translation invariance in the $x_3$ direction constrains the $K$ matrix to be a periodic matrix with a periodicity of $L$. Such a $K$ matrix looks like
\ie\label{eq:Kmatrix_pic}
K=\left(\vcenter{\hbox{
    \begin{tikzpicture}[scale=1]
        \fill[cyan!25, draw=none] (-0.4, 0.4) rectangle (0.4, -0.4);
        \fill[cyan!25, draw=none] (-0.4, 0.4) rectangle (-1.2, 1.2);
        \fill[cyan!25, draw=none] (0.4, -0.4) rectangle (1.2, -1.2);
        \fill[magenta!25, draw=none] (0.4, 0.4) rectangle (1.2, -0.4);
        \fill[magenta!25, draw=none] (-0.4, 1.2) rectangle (0.4, 0.4);
        \fill[magenta!25, draw=none] (-1.2, 0.4) rectangle (-0.4, -0.4);
        \fill[magenta!25, draw=none] (-0.4, -0.4) rectangle (0.4, -1.2);
        \fill[gray!22, draw=none] (-1.2, -0.4) rectangle (-0.4, -1.2);
        \fill[gray!22, draw=none] (0.4, 1.2) rectangle (1.2, 0.4);
        \node at (0.8,0.8) {$M_2$};
        \node at (-0.8,-0.8) {$M_2^T$};
        \node at (1.6, -1.5) {$\ddots$};
        \node at (-1.6, 1.7) {$\ddots$};
        \node at (0,0) {$M_0$};
        \node at (-0.8,0.8) {$M_0$};
        \node at (0.8,-0.8) {$M_0$};
        \node at (0.8,0) {$M_1$};
        \node at (0,0.8) {$M_1$};
        \node at (-0.8,0) {$M_1^T$};
        \node at (0,-0.8) {$M_1^T$};
        \draw[<->] (1.4, 0.38) -- (1.4, -0.38) node[midway, right] {$L$};
    \end{tikzpicture}}}\, \right)~,
\fe
where $M_k$ are $L\times L$ matrices defined in \eqref{eq:Kmatrix}. In order to avoid long-range interactions, $K_{IL+a,JL+b}$ should vanish for sufficiently large $|I-J|$. Mathematically, such matrix is a special case of block-Toeplitz matrices, which have been studied extensively in the literature (see \cite{Toeplitz,deift2012toeplitzmatricestoeplitzdeterminants} for an exposition on Toeplitz matrices and their applications).

\begin{figure*}
    \centering
    \begin{subfigure}[b]{0.31\textwidth}
    \includegraphics[width=\textwidth]{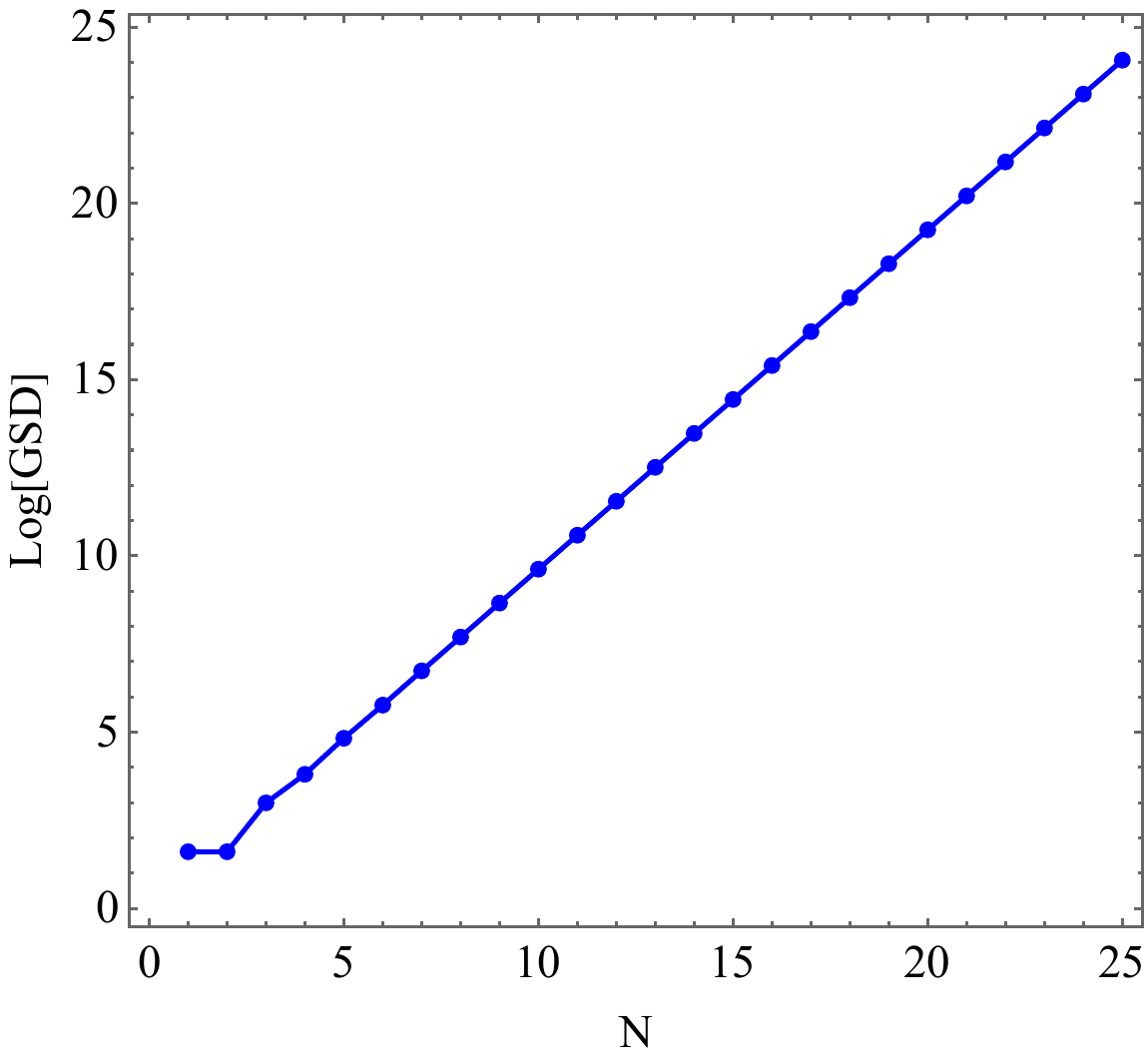}
    \caption{$(3,1)$}\label{fig:31}
    \end{subfigure}\qquad\qquad
    \begin{subfigure}[b]{0.31\textwidth}
    \includegraphics[width=\textwidth]{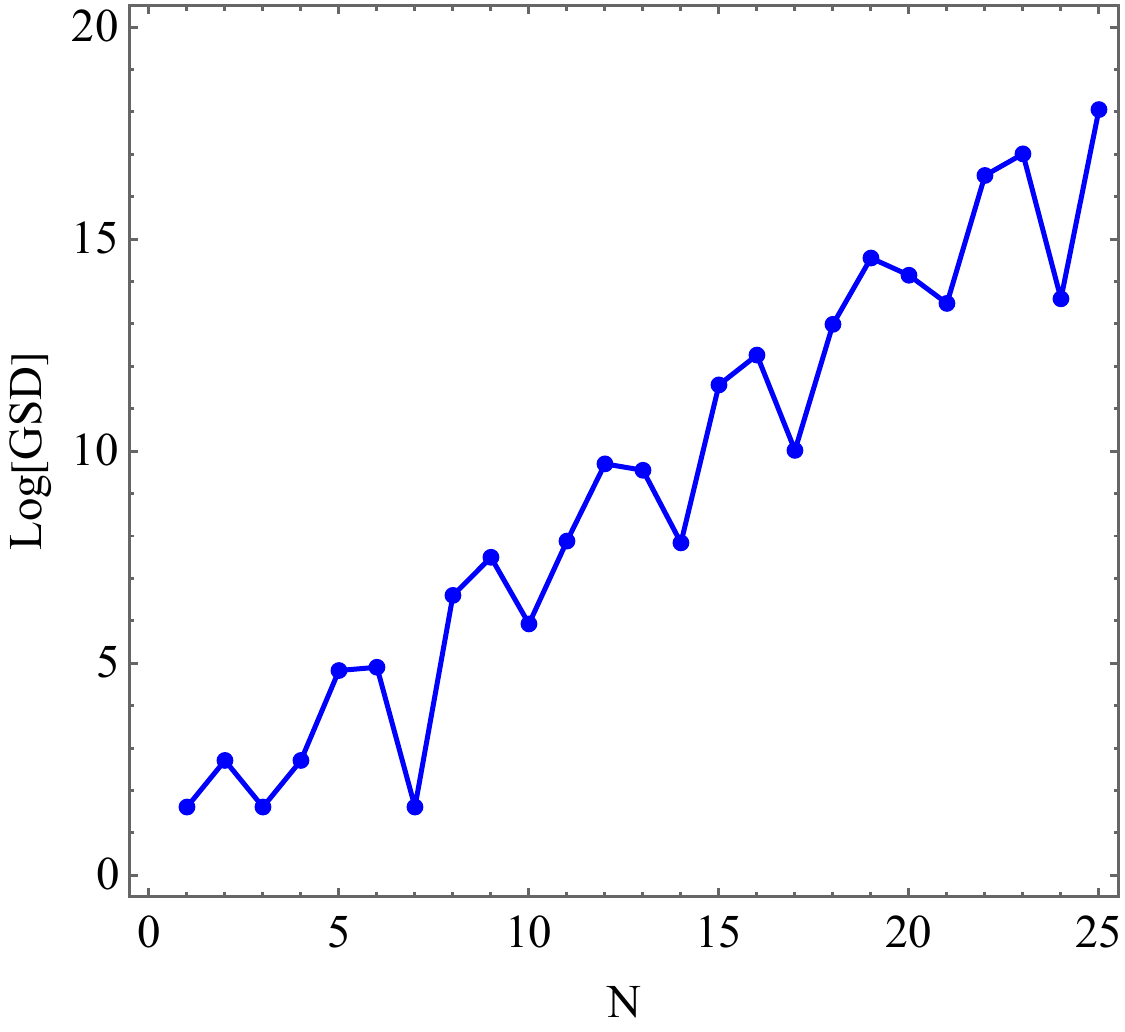}
    \caption{$(1,2)$}\label{fig:12}
    \end{subfigure}\\
    \begin{subfigure}[b]{0.31\textwidth}
    \includegraphics[width=\textwidth]{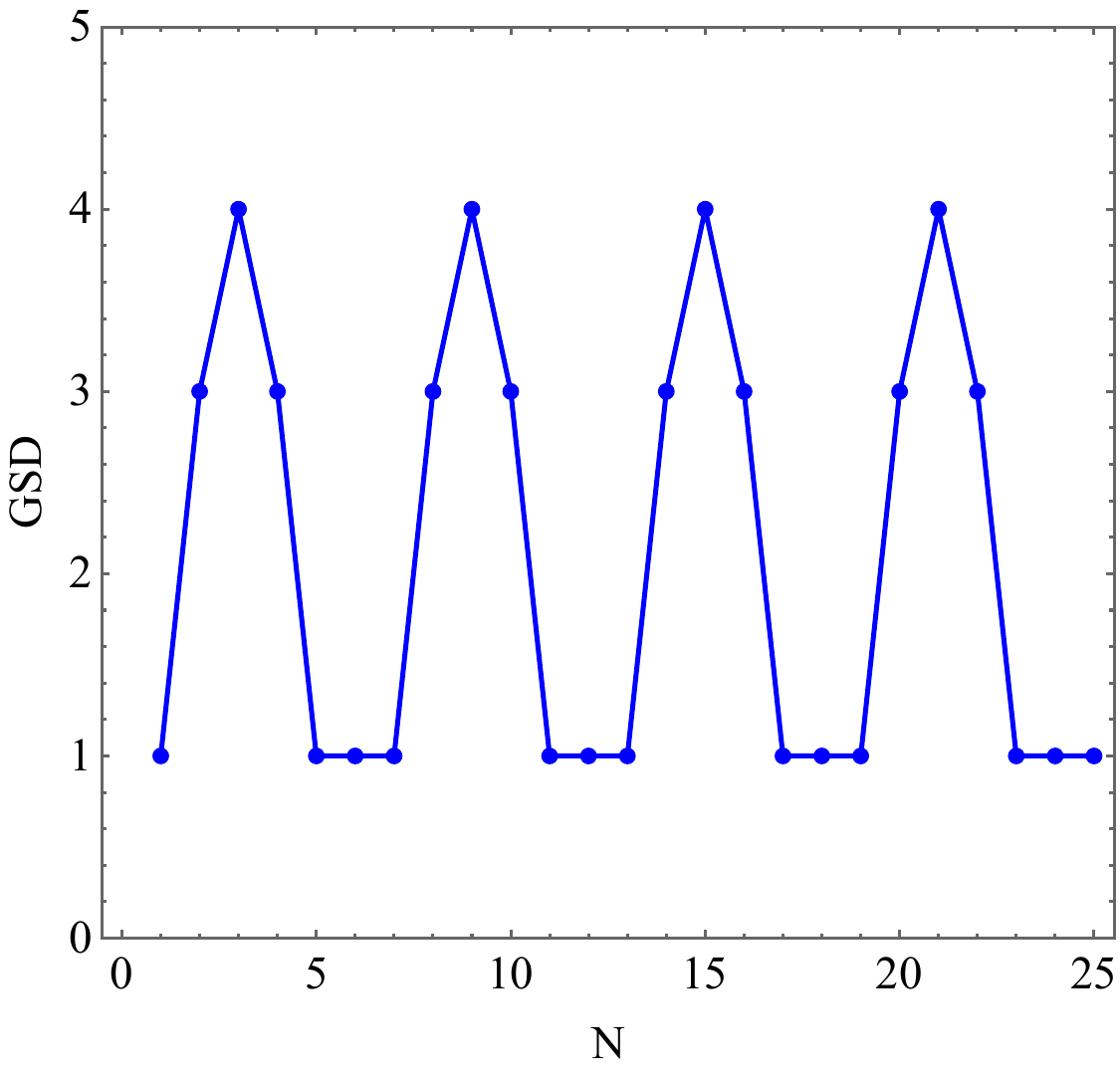}
    \caption{$(-1,1)$}\label{fig:-11}
    \end{subfigure}\qquad\qquad
    \begin{subfigure}[b]{0.31\textwidth}
    \includegraphics[width=\textwidth]{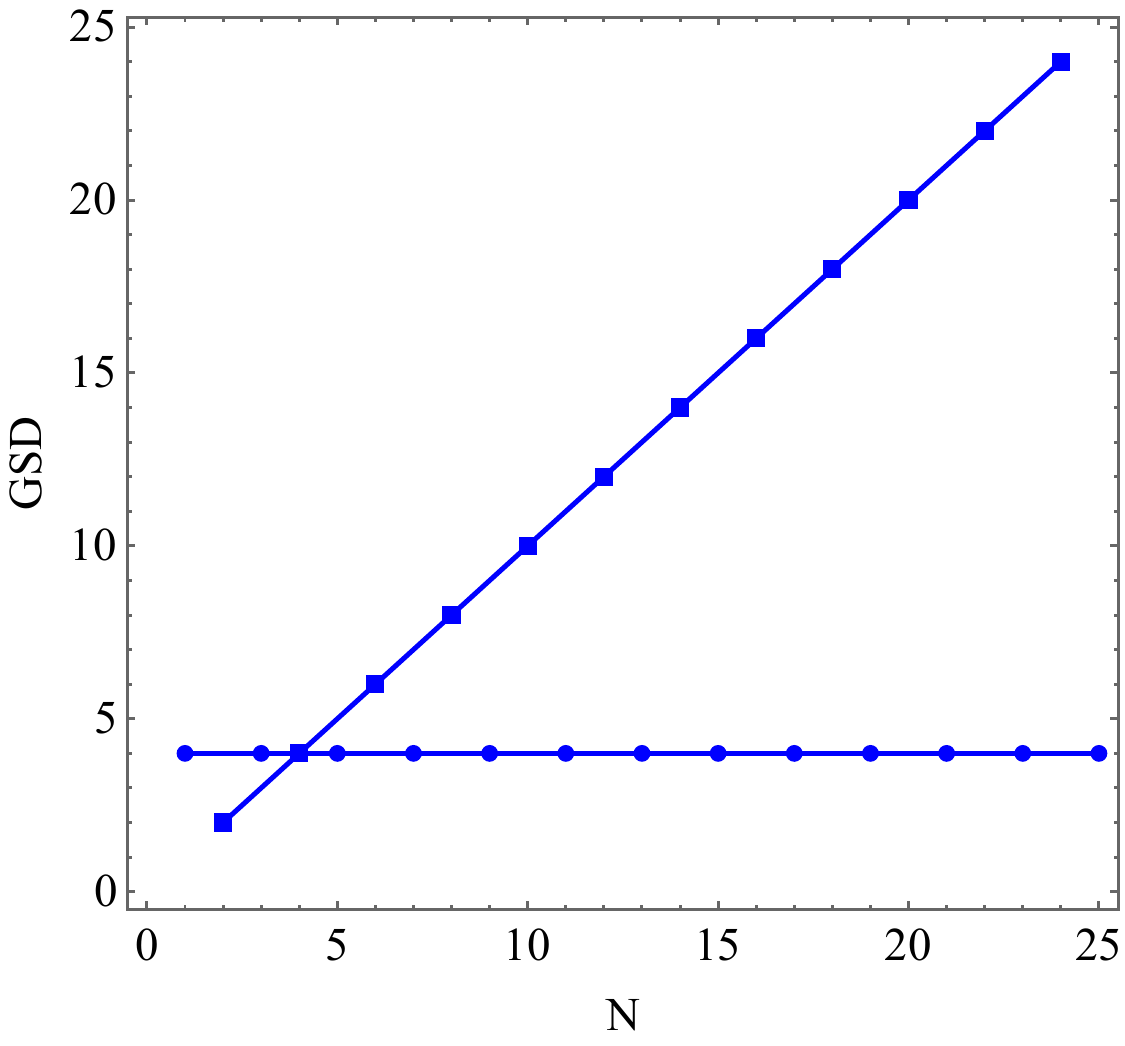}
    \caption{$(2,1)$}\label{fig:21}
    \end{subfigure}
    \captionsetup{justification=Justified}
    \caption{GSD of various tridiagonal iCS theories, labeled by their $(M_0,M_1)$, as a function of the number of layers $N$. Note that the $y$ axis is logarithmic in subfigures~(a) and (b) but not in (c) or (d). In (a), the GSD is not exactly an exponential of $N$ but grows exponentially with $N$ for large $N$. In (b), the GSD fluctuates erractically with an exponentially growing upper bound. Numerics suggests that there is also a lower bound with the same base. In (c), the GSD oscillates as a function of $N$. In (d), the GSD remains a constant for odd $N$ and grows linearly with $N$ for even $N$ }
    \label{fig:GSD_iCS}
\end{figure*}

iCS theories exhibit many unconventional properties such as irrational braiding statistics \cite{PhysRevB.40.11943}, gapless surfaces states \cite{Naud_2000,Naud:2000xa}, exotic global symmetries \cite{Sullivan:2021rbk,Chen:2022hbz} and etc. They generate a plethora of gapped fracton orders as well as new types of robust deconfined gapless fracton orders \cite{Sullivan:2021rbk,Chen:2022hbz}. Some of them can emerge at low energy from three-dimensional fractional quantum Hall systems consisting of stacked two-dimensional electron gases in a perpendicular magnetic field \cite{PhysRevB.40.11943,PhysRevB.42.1339}.

Given the vast zoo of iCS theories, it is natural to seek a systematic approach to organize them. This task, while simpler, is akin to the more challenging endeavor of classifying all fracton orders. At the first level, the iCS theories can be divided into gapped and gapless theories. Among the gapped ones, we can further classify them into foliated and non-foliated fracton orders as defined in \cite{Shirley:2017suz,Shirley_2019_entanglement,Shirley_2019_excitation}.  A foliated fracton order is defined as follows:
\begin{definition}\label{def:foliated}
A fracton model is in a \emph{foliated fracton order} if the model with a larger system size can be mapped to  itself with a smaller system size together with decoupled layers of two-dimensional topological orders using a finite depth local unitary circuit. Otherwise, it is a \emph{non-foliated fracton order}.
\end{definition}
\noindent 
The renormalizability of foliated fracton models is an essential feature that led to the meaningful definition of the foliated fracton phases. Many known Type-I fracton models, including the X-cube model \cite{Vijay:2016phm}, fall into the class of foliated fracton orders.\footnote{In \cite{Vijay:2016phm}, fracton models are divided into two types:~Type-I and Type-II.  In Type-I fracton models, e.g.~the X-cube model \cite{Vijay:2016phm} and the Ising cage-net model \cite{Prem:2018jsn}, the composites of fractons form particles that are mobile along some sub-manifolds. On the other hand, in Type-II fracton models, e.g.~the Haah code \cite{Haah:2011drr}, there are no mobile particles. It is worth emphasizing the notion of Type-I fracton models is not identical to the notion of foliated fracton orders.} The iCS theories with diagonal $K$ matrices also exhibit foliated fracton orders. However, these are not the only examples of foliated iCS theories; for instance, consider the ones presented in \eqref{eq:period2matrix} and \eqref{eq:Kmatrix_test}. 

Non-foliated fracton orders, by contrast, are more exotic and lack a straightforward renormalization picture. These include all Type-II fracton models as well as certain Type-I fracton models, such as the Ising cage-net model \cite{Prem:2018jsn} and various iCS theories \cite{Ma:2020svo}. The existence of fracton models beyond the foliation framework raises key questions: Are these fracton phases are well-defined in the thermodynamic limit, and in what ways can they be considered ‘renormalizable’? For recent developments in the renormalization of non-foliated fracton orders, see \cite{wang2022renormalization}. 

In this paper, we develop tools to determine whether an iCS theory is foliated or non-foliated by analyzing the behavior of its GSD as a function of the linear system size. This leads us to a necessary condition for the theory to be foliated, stated in Proposition \ref{prop:foliation}. Applying this condition, we prove that a large class of iCS theories are non-foliated, opening up new avenues for the investigations of non-foliated fracton orders.

Specifically, we compactify the $x_1,x_2$ direction of the iCS theories on a torus, restrict the number of layers to a finite number $N$ and impose periodic boundary condition in the $x_3$ direction. The index $I$ then runs from $1$ to $N$, and the $K$ matrix becomes an $NL$-dimensional matrix. We find that the GSD of these iCS theories exhibits a very rich pattern as $N$ varies. See the examples in Table \ref{tab:GSD} and Figure \ref{fig:GSD_iCS}. To analyze these patterns, we derive analytic formulae for the GSD in \eqref{eq:det} and \eqref{eq:GSD_gapless_formula}, connecting the different behaviors of GSD to the roots of the determinant polynomial associated with the block-Toeplitz $K$ matrix, defined in \eqref{eq:Laurent_poly}. This allows us to systematically classify the GSD behavior. Furthermore, since the roots of the determinant polynomial also control the spectrum of the iCS theories, we establish a connection between the various GSD behaviors and the spectrum of the iCS theories, as summarized in Table \ref{tab:GSD_iCS2}. 

Mathematically, the GSD is given by the product of the non-zero invariant factors of the Smith normal form of the $K$ matrix. When the $K$ matrix is non-degenerate, it is equivalent to the determinant of the $K$ matrix, and our result \eqref{eq:det} is consistent with the theory of Toeplitz determinants. However, for the degenerate case, the GSD is not related to the determinant of the $K$ matrix. In this case, we derive an alternative formula \eqref{eq:GSD_gapless_formula} for the GSD using a perturbative method.

The rest of the paper is organized as follows. In Section \ref{sec:poly}, we collect some useful mathematical results related to block-Toeplitz symmetric integer matrices. In Section \ref{sec:review}, we review the plane wave spectrum and the braiding statistics in iCS theories. Section \ref{sec:GSD} contains the main results of this paper which relates the behaviors of the GSD of iCS theories with the roots of the determinant polynomial of the black-Toeplitz $K$ matrix. In Section \ref{sec:foliation}, we formulate a necessary condition for the iCS theory to have foliated fracton order. Appendix \ref{app:math} provides more mathematical details regarding polynomial rings and cyclotomic polynomials and proves the two propositions stated in Section \ref{sec:review}. Appendix \ref{app:one-form sym} discusses the structure of the fusion group of an iCS theory when its period is one.

\section{Block-Toeplitz $K$ Matrix}\label{sec:poly}

In this section, we collect some mathematical results regarding block-Toeplitz matrices that are integral and symmetric. We  will refer to this type of matrix as a block-Toeplitz $K$ matrix.

A block-Toeplitz matrix of period $L$ is a block-diagonal-constant matrix of the form
\ie\label{eq:Kmatrix}
K_{IL+a,(I+k)L+b}=M_{k,ab}~.
\fe
For the matrix to qualify as an integer symmetric $K$ matrix, the sub-matrices $M_k$ must be integral and satisfy the condition $M_{-k} = (M_k)^T$. When the $K$ matrix is finite, we always take its size to be $NL$, a multiple of the period $L$, and impose periodic boundary conditions on the indices. Consequently, the $K$ matrix belongs to a special class of block-Toeplitz matrices known as block-circulant matrices.

We can map the $K$ matrix to an $L$-dimensional matrix 
\ie\label{eq:P(u)}
P(u)=\sum_k M_ku^k~,
\fe
whose entries are integer-coefficient Laurent polynomials of a complex variable $u$. The symmetric property of the $K$ matrix implies that $P(u^{-1})=P(u)^T$. On the unit circle $u=\exp(iq)$, $P[\exp(iq)]$ is essentially the Fourier transform of the $K$ matrix, and it is often referred to as the symbol of the $K$ matrix in the literature of Toeplitz matrix.

The eigenvectors of the $K$ matrix take the form of
\ie\label{eq:eigenvec}
v_{IL+a}(q)=v_a(q)e^{iq I}~,
\fe
where $v_a(q)$ is an eigenvector of $P[\exp(iq)]$ and $q\sim q+2\pi$. For a given, $q$ there are $L$ independent eigenvectors. We denote their eigenvalues by $\lambda_a(q)$ where $a=1,...,L$. When the size of the $K$ matrix -- $NL$ is finite, $q$ are quantized to be $2\pi k/N$ with integer $k$ and in particular can only be a rational multiple of $2\pi$. The quantization condition is removed when $N\to\infty$.

The determinant of $P(u)$ is also a Laurent polynomial
\ie\label{eq:Laurent_poly}
D(u)=\det[P(u)]~,
\fe 
which will play a central role in our calculations of the GSD. We refer to $D(u)$ as the \textit{determinant polynomial} of the $K$ matrix. We can factorize $D(u)$ as
\ie\label{eq:factorize}
D(u)=C u^{-\xi}\prod_{\alpha}(u-u_\alpha)^{\Gamma_\alpha}
\fe
where $u_\alpha$ are the distinct roots of $D(u)$, $\Gamma_\alpha$ are their multiplicities and
\ie\label{eq:def_degree}
\xi=\frac{1}{2}\sum_{\alpha}\Gamma_\alpha
\fe
is the power of the leading order term in $D(u)$. Because of the symmetric property of the $K$ matrix, $D(u)=D(u^{-1})$ and the roots of $D(u)$ are distributed such that they are invariant under the map $u\mapsto u^{-1}$.

The roots on the unit circle are particularly special because they are related to the zero eigenvalues of $K$. On the unit circle, we have
\ie\label{eq:eigenvalue_polynomial}
D(e^{iq})=\prod_{a=1}^L\lambda_a(q)~.
\fe
Thus, when $D(u)$ has a root $u_\alpha=\exp(iq_\alpha)$ on the unit circle, there is at least one eigenvalue vanishing at $q=q_\alpha$ and vice versa. In principle, there could be more than one zero eigenvalue. We will refer to these roots on the unit circle as \emph{unit roots} and those away from the unit circle as \emph{non-unit roots}.

Among the unit roots, those whose phases are rational multiples of $2\pi$ are even more special, and we refer to them as \emph{rational roots}; unit roots that are not rational are referred to as \emph{irrational roots}. There are two useful mathematical results concerning the distributions of the rational roots, which we state here and prove in  Appendix~\ref{app:math} using the theory of polynomial rings:
\begin{proposition}\label{prop:primitive_roots}
    An $m$th root of unity $\exp(2\pi i k/m)$ is called \textit{primitive} if $k$ is coprime with $m$. If $D(u)$ has a primitive $m$th root of unity as one of its roots, then it has all the other primitive $m$th roots of unity as its roots with the same multiplicity.
\end{proposition}
\begin{proposition}\label{prop:all_rational}
    If $D(u)$ has leading coefficient $C=1$ and has only unit roots, then all of its roots are rational.
\end{proposition}

\section{Review of iCS Theories}\label{sec:review}

In this section, we review the plane wave spectrum \cite{Ma:2020svo,Sullivan:2021rbk,Chen:2022hbz} and the braiding statistics \cite{Ma:2020svo,Chen:2022hbz} in iCS theories. In particular, we highlight how different types of roots of the determinant polynomial affect these physical observables. In Section \ref{sec:GSD}, we will again see that different types of roots have drastically different effects on the behavior of the GSD as a function of the number of layers $N$.

\subsection{Plane Wave Spectrum}
An iCS theory has propagating plane waves 
\ie
\mathcal{A}_\mu^{I,a}=\mathcal{C}_{\mu}v_a(q)e^{i\omega t+ik_x+ik_y+iqI}~,
\fe
with dispersion relations 
\ie
\omega^2=k_x^2+k_y^2+\frac{g^4}{4\pi^2}\lambda(q)^2~.
\fe
Here $q\sim q+2\pi$ is the momentum in the $x_3$ direction and $v(q)$ is an eigenvector of $P[\exp(iq)]$ with eigenvalue $\lambda(q)$. 

The theory is gapless at momentum $q$ if $\lambda_a(q)=0$ for any $a=1,...,L$, or equivalently if $D(u)$ has a root on the unit circle at $u=\exp({iq})$. Below and in Table \ref{tab:GSD_iCS2}, we summarize the effect of the three types of roots on the plane wave spectrum:
\begin{itemize}[leftmargin=*]
    \item 
    A non-unit root does not lead to any gapless mode.
    \item
    An irrational root leads to gapless modes in the $N\to\infty$ limit but not when $N$ is finite. It is because when $N$ is finite, the momenta $q$ are quantized and can only be rational multiples of $2\pi$. As $N\to\infty$, $q$ can take finer and finer values and eventually the gap closes at the irrational momentum associated to the irrational root.
    \item 
    A rational root at a primitive $m$th root of unity $u=\exp(2\pi ik/m)$ implies that the spectrum closes at $q=2\pi k/m$ when $N$ is a multiple of $m$. Because of Proposition \ref{prop:primitive_roots}, when the spectrum closes at $q=2\pi k/m$, it necessarily closes at $q={2\pi k'}/{m}$ for all other $k'$ coprime to $m$.
\end{itemize}

\subsection{Braiding Statistics}

We can take a charge at layer $I$ and braid it with another charge at layer $I+z$. This induces a braiding phase $\exp[i\Phi(z)]$. In the $N\to\infty$ limit, it was shown that as the size of the braiding trajectory approaches infinity, $\Phi(z)$ converges and depends only topologically on the braiding trajectory if the theory is gapped or gapless with linear dispersion around the gapless points \cite{Chen:2022hbz}. In contrast, if there exists a gapless point with a higher order dispersion, $\Phi(z)$ diverges as the size of the braiding trajectory approaches infinity. Below and in Table \ref{tab:GSD_iCS2}, we focus on the former situation so that $\Phi(z)$ has a convergent asymptotic value and summarize the effect of the three types of roots on the asymptotic braiding phase:
\begin{itemize}[leftmargin=*]
    \item 
    A non-unit root $u_\alpha$ contributes to the asymptotic braiding phase an exponentially decaying term proportional to $|u_\alpha|^{\pm |z|}$. The sign of the exponent is chosen such that this term decays with $|z|$.
    \item 
    A unit root $u_\alpha=\exp(iq_\alpha)$, including both rational and irrational root, contributes to the asymptotic braiding phase an oscillatory term proportional to $\sin(q_\alpha z)$.
\end{itemize}

\begin{table*}
    \centering
    \begin{tabular}{|c|c|c|c|c|}
        \hline
        Polynomial & Effect on GSD & Spectrum 
        \\
        \hline
        Leading Coefficient & An exponential factor & $-$ 
        \\
        \hline
        Non-unit root & Asymptotically (but never strictly) exponential & Gapped mode 
        \\
        \hline
        Irrational root & Erratic oscillation with exponential envelope  & Gapless mode in infinite $N$ 
        \\
        \hline
        Rational root & Branch-wise constant or polynomial & Gapless mode in finite $N$ 
        \\
        \hline
    \end{tabular}
    \caption{The effect of the leading coefficient $C$ and the roots $u_\alpha$ of the determinant polynomial $D(u)$ in \eqref{eq:factorize} on the GSD and the spectrum.}
    \label{tab:GSD_iCS2}
\end{table*}

\section{Ground State Degeneracy}
\label{sec:GSD}
The GSD of an iCS theory can be inferred from its one-form global symmetry \cite{Gaiotto_2015}. Suppose the Smith normal form of the $K$ matrix is
\ie\label{eq:SNF}
VKW=R=\left(\begin{array}{cccccc}
     r_1&  \\
     & r_2 \\
     & & \ddots\\
     & & & r_{N}
\end{array}\right)~,
\fe
where $V$ and $W$ are $GL(NL,\mathbb{Z})$ matrices, and $r_i$ are the invariant factors of the $K$ matrix, which are integers that divide the succeeding ones $r_{i+1}$. Let $N_0$ and $N_1$ be the number of zero and non-zero eigenvalues of the $K$ matrix, respectively. Then the first $N_1$ of the $r_i$'s are non-zero, whereas the last $N_0$ of them are zero. The one-form global symmetry is generated by the transformation
\ie
\mathcal{A}_\mu^{I,a}\rightarrow \mathcal{A}_\mu^{I,a}+\Lambda_\mu^{I,a}~,\quad \Lambda_\mu^{I,a}=\sum_{i=1}^{NL}  W_{IL+a,i} \lambda_\mu^i~,
\fe
where $\lambda_\mu^i$ are $\mathbb{Z}_{r_i}$ gauge field for $1\leq i\leq N_1$ and flat $U(1)$ gauge fields for $N_1<i\leq NL$ \cite{Chen:2022hbz}. This transformation shifts the Lagrangian \eqref{eq:action} by 
\ie
\mathcal{L}\rightarrow\mathcal{L}&+\frac{i}{2\pi} \epsilon^{\mu\nu\rho} K_{IL+a,JL+b} \Lambda_\mu^{I,a}\partial_\nu \mathcal{A}^{J,b}_\rho
\\
&+\frac{i}{4\pi} \epsilon^{\mu\nu\rho}K_{IL+a,JL+b}\Lambda_\mu^{I,a}\partial_\nu \Lambda_\rho^{J,b}~.
\fe
Although the shift of the Lagrangian is non-trivial, it integrates to an integer multiple of $2\pi i$ as 
\ie
\sum_I K_{IL+a,JL+b} \Lambda_\mu^{I,a}
\fe
is a flat gauge field with $2\pi\mathbb{Z}$ holonomy. The partition function is therefore invariant under the transformation. In summary, the theory has a 
\ie
U(1)^{N_0}\times\prod_{i=1}^{N_1}\mathbb{Z}_{r_i}
\fe
one-form global symmetry, or relatedly a $\prod_{i=1}^{N}\mathbb{Z}_{r_i}$ fusion group with $\mathbb{Z}_0$ identified as a trivial group.

In Appendix~\ref{app:one-form sym}, we show that when the $K$ matrix has period one and $\text{gcd}\{M_k\}=1$, all of the $r_i$ are equal to one except for the last $2\xi$ of them, where $\xi$ is defined in \eqref{eq:def_degree}. In this case, the fusion group of the fractional excitations is
\ie
G=\mathbb{Z}_{r_{N-2\xi+1}}\times\cdots\times\mathbb{Z}_{r_N}~,
\fe
which has a finite number of large cyclic components for large $N$. Here, a group such as $\mathbb{Z}_6=\mathbb{Z}_2\times\mathbb{Z}_3$ is viewed as having one cyclic component instead of two.

Now, consider compactifying the $x_1,x_2$ direction of the iCS theory to a two-dimensional compact manifold. This leads to a discrete spectrum that is equally degenerate at every energy level. This degeneracy in particular is the same as the GSD. It is worth emphasizing that the theory may or may not be gapless when the volume of the compact manifold approaches infinity, but this does not affect the well-definedness and the value of the GSD. 

The non-trivial GSD is a consequence of the spontaneous breaking of the discrete one-form symmetry.\footnote{In contrast, a $U(1)$ one-form symmetry cannot be spontaneously broken in two spatial dimensions \cite{Gaiotto_2015,Lake:2018dqm}.} It depends only on the topology of the two-dimensional manifold. When the manifold is a torus, the GSD is given by the order of the discrete one-form symmetry group. On a more general genus $g$ manifold, 
\ie\label{eq:GSD_SNF}
\text{GSD}=\left(\prod_{i=1}^{N_1} r_i\right)^g~.
\fe
Although the GSD depends only topologically on the manifold in the $x_1,x_2$ direction, it can have a non-trivial dependence on the size of the $x_3$ direction, \textit{i.e.}\ the number of layers $N$, as in other fracton orders. Our main focus in the next two subsections is to compute the GSD as a function of $N$. Without loss of generality, we will restrict the manifold in the $x_1,x_2$ direction to be a torus in the rest of the paper.

\subsection{Non-degenerate $K$ Matrix}
Let us first compute the GSD when the $K$ matrix is non-degenerate \textit{i.e.}\ has no zero eigenvalues. In this case, the GSD reduces to the absolute value of the determinant of the $K$ matrix (see for example the review \cite{Wen:1995qn})
\ie
\text{GSD}=\prod_{i=1}^{NL}r_i=\text{det}(V^{-1}KW^{-1})=|\text{det}(K)|~,
\fe
where we used the fact that every $GL(NL,\mathbb{Z})$ matrix $V$ has $|\det V|=1$. Determinant of the $K$ matrix is given by the product of the eigenvalues. Using the relation \eqref{eq:eigenvalue_polynomial} between the eigenvalues and the determinant polynomial, we obtain
\ie
\text{GSD}=\left|\prod_{k=1}^N\prod_{a=1}^L\lambda_a\left(\frac{2\pi k}{N}\right)\right|=\left|\prod_{k=1}^N D(e^{{2\pi ik}/{N}})\right|~.
\fe
Substituting \eqref{eq:factorize} into the above formula, the GSD can then be expressed in terms of $u_\alpha$, the roots of $D(u)$
\ie\label{eq:det}
\text{GSD}
&=|C|^N\prod_\alpha\left|\prod_{k=1}^N(e^{{2\pi ik}/{N}}-u_\alpha)\right|^{\Gamma_\alpha}
\\
&=|C|^N\prod_\alpha\left|u_\alpha^N-1\right|^{\Gamma_\alpha}~.
\fe
In the last equality, we used the fact that $u^N-1$ has $N$ distinct roots at $\exp(2\pi ik/N)$ with $k=1,...,N$ and thus can be factorized into $\prod_{k=1}^N[u-\exp(2\pi ik/N)]$. The overall coefficient of the factorization is fixed by comparing the coefficient of $u^N$ with the expansion of the factorization. By expanding \eqref{eq:det}, we learn that the GSD for non-degenerate $K$ matrices can be expressed as a sum of exponentials of $N$.

Mathematically, \eqref{eq:det} computes the Toeplitz determinant of the block-Toeplitz $K$ matrix, whose asymptotic behaviors have been a subject of considerable interest in the mathematical literature, with connections to Szegö’s Theorem \cite{szego} and the Fisher-Hartwig conjecture \cite{FHconjecture}. The expression in \eqref{eq:det} can be interpreted as the absolute value of the \textit{resultant} of the polynomials $u^\xi D(u)$ and $u^N-1$. We refer to \cite{cohn2000classic} for more details on resultant.

\subsection{Degenerate $K$ Matrix}

Next, we move on to consider degenerate $K$ matrices. Because there are zero eigenvalues, the GSD is no longer simply given by $|\det(K)|$. Nevertheless, we can still determine how it depends on $N$. We will first state the result and then derive it.

The main result is as follows: Let $u_\alpha$ be the {distinct} roots of the determinant polynomial $D(u)$, $\Gamma_{\alpha}$ be the multiplicity of $u_\alpha$, and  $\Delta_\alpha$ be the kernel dimension of $P(u_\alpha)$ in \eqref{eq:P(u)}. $\Delta_\alpha$ is always upper bounded by $\Gamma_\alpha$.\footnote{To understand the difference between $\Gamma_\alpha$ and $\Delta_\alpha$ better, it is helpful to draw an analogy with the \textit{algebraic multiplicity} $\Gamma$ and the \textit{geometric multiplicity} $\Delta$ of an eigenvalue $\lambda$ in linear algebra. In linear algebra, the polynomial of interest is the characteristic polynomial of a matrix $K$, the algebraic multiplicity is the number of roots of the characteristic polynomial at the given eigenvalue $\lambda$ and the geometric multiplicity is the kernel dimension of $K-\lambda \mathbb{I}$. The quantity $\Gamma-\Delta$ intuitively measures the obstacle to diagonalize the matrix. }
Define also an index set $\mathcal{I}=\left\{\alpha:u_\alpha^N=1\right\}$, the set of roots of $D(u)$ which correspond to gapless modes when the number of layer is $N$. The set $\mathcal{I}$ generally depends on $N$, and when discussing the dependence of the GSD on $N$, we only consider those values of $N$ that yield the same $\mathcal{I}$. For example, in the context of $D(u)=u-1+u^{-1}$, the roots are $u=\exp(\pm 2\pi i/6)$ and thus we consider those $N$ that are divisible by 6 separately from the rest. For the values of $N$ that share the same index set $\mathcal{I}$, the GSD takes the form
\begin{equation}\label{eq:GSD_gapless_formula}
    \text{GSD}=A|C|^N N^{\sum_{\alpha\in \mathcal{I}}(\Gamma_\alpha - \Delta_\alpha)} \prod_{\beta\notin \mathcal{I}}|u_\beta^N-1|^{\Gamma_\beta}~,
\end{equation}
where $A$ is an $N$-independent constant.\footnote{The precise value of $A$ is a complicated expression involving basic algebraic number theory, and we will not determine it in this paper. Practically, $A$ is most conveniently fixed by fitting \eqref{eq:GSD_gapless_formula} for a small $N$.} The dependence of the GSD on $N$ has an additional polynomial factor besides the usual terms in \eqref{eq:det}. This is controlled by the roots in $\mathcal{I}$. Mathematically, \eqref{eq:GSD_gapless_formula} computes the product of the non-zero invariant factors of the block-Toeplitz $K$ matrix.

We now derive the formula \eqref{eq:GSD_gapless_formula} starting from \eqref{eq:GSD_SNF}. In order to evaluate \eqref{eq:GSD_SNF}, we add a small perturbation $\epsilon$ (times the identity matrix) to $K$ and turn it into $K+\epsilon\mathbb{I}$. Since $K+\epsilon\mathbb{I}$ is not integral, it does not physically correspond to a gauge invariant iCS theory. Instead, this perturbation is merely a mathematical technique. Now take $V,W$ as in \eqref{eq:SNF} and consider
\begin{align}
    V(K+\epsilon\mathbb{I})W&=
\begin{pmatrix}
    \left[\hat R+\mathcal{O}(\epsilon)\right]_{N_1\times N_1} & \mathcal{O}(\epsilon)_{N_1\times N_0}\\
    \mathcal{O}(\epsilon)_{N_0\times N_1} & \epsilon\, H_{N_0\times N_0}
\end{pmatrix} \label{eq:CS_gapless_block} \nonumber
\end{align}
Here $\hat R=\text{diag}(r_1,r_2,...,r_{N_1})$ and $H$ is an $N_0\times N_0$ matrix with order 1 coefficients.
To leading order in $\epsilon$,
\begin{align*}
    \det(K+\epsilon\mathbb{I})&=\det(V(K+\epsilon)W)\\
    &=\epsilon^{N_0}\det(H)\det(\hat{R})~,
\end{align*}
and hence
\begin{equation}\label{eq:GSD_gapless_key}
    \text{GSD}=|\det(\hat{R})|=\left|\frac{\det(K+\epsilon)}{\epsilon^{N_0}\det(H)}\right|.
\end{equation}
This is the key equation for calculating the GSD. 

Because the numerator is the determinant of the \textit{non-degenerate} block-Toeplitz matrix $K+\epsilon\mathbb{I}$, we can calculate it using \eqref{eq:det}, which relates the determinant with the roots of $\det[P(u)+\epsilon\mathbb{I}]$.
For small $\epsilon$, the roots of $\det[P(u)+\epsilon\mathbb{I}]$ are close to the roots of $\det[P(u)]$. To find the new roots, we expanding $\det[P(u)+\epsilon\mathbb{I}]$ in the vicinity of $u_\alpha$
\begin{equation}\label{eq:det_expand_general}
\det[P(u)+\epsilon\mathbb{I}]=\det[P(u_\alpha)+\epsilon\mathbb{I}]+\eta_\alpha (u-u_\alpha)^{\Gamma_\alpha}+\cdots,
\end{equation}
where $\eta_\alpha\neq0$ and ``$\cdots$'' stands for higher order terms in $u-u_\alpha$. Let $\lambda_a$ be the eigenvalues of $P(u_\alpha)$ with the first $\Delta_\alpha$ of them vanishing.
Then
\ie
\det[P(u_\alpha)+\epsilon\mathbb{I}]=\prod_{a=1}^L (\lambda_a+\epsilon)=b_\alpha\epsilon^{\Delta_\alpha}+\cdots~.
\fe
where
$b_{\alpha}=\prod_{i=\Delta_\alpha+1}^L \lambda_a\neq0$. 
Thus \eqref{eq:det_expand_general} becomes
\ie
\det[P(u)+\epsilon\mathbb{I}]=\eta_\alpha (u-u_\alpha)^{\Gamma_\alpha}+b_{\alpha}\epsilon^{\Delta_\alpha}+\cdots~,
\fe
and the new roots are 
\ie
u_{\alpha,m}(\epsilon)=u_\alpha+\left(B_\alpha \epsilon^{\Delta_\alpha}\right)^{1/\Gamma_\alpha}e^{2\pi im/\Gamma_\alpha}~.
\fe
Here $m=1,\ldots,\Gamma_\alpha$ labels the new roots and $B_\alpha=-b_{\alpha}/\eta_\alpha$. To avoid ambiguity, the $\Gamma_\alpha$th root of $B_\alpha \epsilon^{\Delta_\alpha}$ is chosen to be a specific one for all $m$. Here we see that the $\Gamma_\alpha$ degenerate roots of $D(u)$ at $u=u_\alpha$ are split into $\Gamma_\alpha$ distinct roots by the perturbation.

We now substitute the new roots into \eqref{eq:det} to calculate $\det[P(u)+\epsilon\mathbb{I}]$. Each new root contributes a factor of $u_{\alpha,m}(\epsilon)^N-1$. If $\alpha\notin \mathcal{I}$, the correction due to the perturbation is insignificant and
\ie
u_{\alpha,m}(\epsilon)^N-1=u_{\alpha}^N-1+\cdots~.
\fe
On the other hand, if $\alpha\in \mathcal{I}$, then $u_\alpha^N=1$ and
\begin{equation}\label{eq:gapless_root_contribution}
    u_{\alpha,m}(\epsilon)^N-1
    =Nu_\alpha^{-1}\left(B_\alpha \epsilon^{\Delta_\alpha}\right)^{1/\Gamma_\alpha}e^{2\pi im/\Gamma_\alpha}+\cdots~.
\end{equation}
Multiplying all the contributions together, the numerator of \eqref{eq:GSD_gapless_key} is found to be
\begin{equation}\label{eq:GSD_gapless_numerator}
    |\det(K+\epsilon)|=|C|^N \prod_{\alpha\in \mathcal{I}}N^{\Gamma_\alpha}|B_\alpha \epsilon^{\Delta_\alpha}|\prod_{\beta\notin \mathcal{I}}|u_\beta^N-1|^{\Gamma_\beta}+\cdots~.
\end{equation}

Next, we discuss the denominator of \eqref{eq:GSD_gapless_key}. Let $\hat W$, $\hat V^T$ be the $NL\times N_0$ matrix consisting of the last $N_0$ columns of $W$, $V^T$ of the Smith normal form \eqref{eq:SNF} respectively. Then
\ie
H=\epsilon^{-1}\hat V^T(K+\epsilon)\hat W=\hat V^T \hat W.
\fe
The columns of $\hat W$, $\hat V^T$ are the null vectors of $K$ with integral coefficients. They can be decomposed into linear combinations of null vectors of the form \eqref{eq:eigenvec} with $q_\alpha=-i\log u_\alpha$ for $\alpha\in\mathcal{I}$. All the null vectors are thus periodic vectors, and each can be constructed by repeating some vector $\mathcal{O}(N)$ times.
It then implies that every entry of $\hat V^T\hat W$ is proportional to $N$ and thus
\begin{equation}\label{eq:GSD_gapless_denominator}
    \det(H)\propto N^{N_0}.
\end{equation}

Finally, using $N_0=\sum_{\alpha\in I}\Delta_\alpha$ to cancel all the $\epsilon$'s and absorbing various constant factors into an overall constant $A$, \eqref{eq:GSD_gapless_numerator} and \eqref{eq:GSD_gapless_denominator} are combined to give \eqref{eq:GSD_gapless_formula}. This concludes our derivation of \eqref{eq:GSD_gapless_formula}.

\subsection{Effects of Three Types of Roots}

Equipped with the formula \eqref{eq:det}, \eqref{eq:GSD_gapless_formula} for GSD, we are now ready to examine how different types of roots affect the GSD. 

\subsubsection{Non-unit Roots}

First, let us discuss the effect of the non-unit roots. Recall that the list of roots are invariant under the map $u\mapsto u^{-1}$. Thus every non-unit root $u_\alpha$ is paired up with another non-unit root $u_{\alpha^*}=u_\alpha^{-1}$ and within a pair, one of the roots has magnitude greater than 1 while the other has magnitude less than one. According to \eqref{eq:det}, a pair of non-unit roots contribute to the GSD as
\ie
|(u_\alpha^N-1)(u_{\alpha^*}^{N}-1)|^{\Gamma_\alpha}\xrightarrow[]{\text{$N\gg1$}}
\begin{dcases}
    |u_\alpha|^{N\Gamma_\alpha}~, \ &|u_\alpha|>1
    \\
    |u_\alpha|^{-N\Gamma_\alpha}~,\ &|u_\alpha|<1
\end{dcases}~,
\fe
which grows exponentially with $N$ for large $N$.

If an iCS theory has only non-unit roots, then half of the roots has magnitude greater than one while the other half has magnitude less than one. When $N$ is large, the asymptotic GSD grows exponentially as
\ie\label{eq:GSD_asym}
\text{GSD}=\left(|C|\prod_{|u_\alpha|>1}|u_\alpha|^{\Gamma_\alpha}\right)^N~.
\fe
This agrees with Szeg\"o's theorem \cite{szego}, which states that when $D(e^{i\theta})$ is a continuous, non-vanishing function on the unit circle $S^1$, the Toeplitz determinant of the $K$ matrix grows exponential as
\ie\label{eq:szego}
\log \text{GSD}=\log \det K=\frac{N}{2\pi}\int_{-\pi}^\pi D(e^{i\theta})d\theta+\mathcal{O}(N)~.
\fe
Indeed, using \eqref{eq:factorize} and express \eqref{eq:szego} as a complex integral, we can recover \eqref{eq:GSD_asym} by the residue theorem.

As an example, consider a theory with a tridiagonal $K$ matrix
\ie\label{eq:tridiagonal}
K=\left(
\begin{array}{ccccccccccc}
\ddots&&&\\
&M_0&M_1&&\\
&M_1&M_0&M_1&\\
&&M_1&M_0&M_1\\
& &&M_1&M_0&\\
&&&&&\ddots
\end{array}
\right)~.
\fe
The theory has $D(u)=M_1u+M_0+M_1u^{-1}$ with roots at
\ie
u_\pm=\frac{M_\pm}{M_1}~,\quad M_\pm=\frac{-M_0\pm\sqrt{M_0^2-4M_1^2}}{2}~.
\fe
The roots are both non-unit roots if $|M_0|>2|M_1|$ and the GSD is
\ie\label{eq:GSD_31}
\text{GSD}=\left|M_+^N+M_-^N-2M_1^N\right|
\fe
In Figure \ref{fig:31}, we plot the GSD as a function of $N$ when $(M_0,M_1)=(3,1)$. It grows exponentially for large $N$.

\subsubsection{Irrational Roots}

Next, we move on to discuss the effect of the irrational roots. An irrational root can be expressed as $u_\alpha=\exp(iq_\alpha)$. Again as non-unit roots, it is paired up with another irrational root $u_{\alpha^*}=\exp(-iq_\alpha)$. According to \eqref{eq:det}, a pair of them contributes to the GSD as
\ie
|(u_\alpha^N-1)(u_{\alpha^*}^{N}-1)|^{\Gamma_\alpha}=\left|2\sin\left(\frac{q_\alpha N}{2}\right)\right|^{2\Gamma_\alpha}~,
\fe
which oscillates as a function of $N$. Since $q_\alpha$ is an irrational multiple of $2\pi$, the oscillation is erratic as $N$ increases by integer steps.

If an iCS theory has irrational roots, its asymptotic GSD no longer grows exponentially with $N$ but rather fluctuates erratically. Suppose there are only non-unit roots and irrational roots, then the asympotic GSD is 
\ie\label{eq:GSD_rational_nonunit}
\text{GSD}=\left(|C|\prod_{|u_\alpha|>1}|u_\alpha|^{\Gamma_\alpha }\right)^N\prod_{\substack{|u_\beta|=1\\u_\beta=e^{iq_\beta}}}\left|2\sin\left(\frac{q_\beta N}{2}\right)\right|^{\Gamma_\beta}~,
\fe 
with an exponentially growing envelop 
\ie
\text{GSD}\leq\left(|C|\prod_{|u_\alpha|>1}|u_\alpha|^{\Gamma_\alpha }\right)^N\prod_{{|u_\beta|=1}}2^{\Gamma_\beta}~.
\fe
It is consistent with Fisher-Hartwig conjecture \cite{FHconjecture}, which in our case implies an upper bound on the GSD:
\ie
\text{GSD}\leq \# N^{\frac{1}{4}\sum_{|u_\beta|=1} \Gamma_\beta^2}\left(|C|\prod_{|u_\alpha|>1}|u_\alpha|^{\Gamma_\alpha }\right)^N~,
\fe
where $\#$ is an $N$-indepedent constant.
Note that the base of the exponentially growing envelop $|C|\prod_{|u_\alpha|>1}|u_\alpha|^{\Gamma_\alpha}$ is always greater than one if there exists an irrational root. If $|C|>1$, the base is necessarily greater than one. An example of this kind is a period one $K$ matrix with $D(u)=2u+1+2u^{-1}$, which has two irrational roots. If $|C|=1$, there must be non-unit roots and thus the base is still greater than one.
It is because otherwise all the roots are unit roots and according to Proposition \ref{prop:all_rational} they must all be rational, which contradicts with the assumption that there exists at least one irrational root. An example of this kind is a period one $K$ matrix with $D(u)=u^2+2u+1+2u^{-1}+u^{-2}$, which has two irrational roots and two non-unit roots.

As an example, consider a theory with a tridiagonal $K$ matrix \eqref{eq:tridiagonal}. The roots are unit roots when $|M_0/M_1|\leq2$. They are generally irrational unless  $|M_0/M_1|=0,\pm1,\pm2$ according to the Niven's theorem \cite{Niven}.
The GSD is
\ie
\text{GSD}=4|M_1|^N\sin^2\left(\frac{qN}{2}\right)~.
\fe
where $q=\arccos(-M_0/2M_1)$.
In Figure \ref{fig:12}, we plot the GSD as a function of $N$ when $(M_0,M_1)=(1,2)$. It exhibits an erratic pattern with an exponentially growing upper bound. Numerically, it seems that the GSD also has an exponentially growing lower bound with the same base. However, we do not know how to prove the lower bound.

\subsubsection{Rational Roots}
Finally, we discuss the effect of the rational roots. A rational root can be expressed as $u_\alpha=\exp(2\pi ik/m)$ where $k$ and $m$ are coprime integers. Such a rational root is a primitive $m$th root of unity. According to proposition \ref{prop:primitive_roots}, all the other primitive $m$th roots of unity are also roots of $D(u)$ with the same multiplicity.

The contribution of these primitive $m$th roots of unity to the GSD depends on whether $N$ is divisible by $m$ or not. If $N$ is not divisible by $m$, according to \eqref{eq:det} the contribution to the GSD is an oscillatory function of $N$
\ie\label{eq:rational_mod_1}
\prod_{\substack{1\leq k\leq m\\\gcd(k,m)=1}}\left|2\sin\left(\frac{ k\pi N}{m}\right)\right|^{\Gamma_\alpha}~.
\fe
If $N$ is divisible by $m$, the $K$ matrix becomes degenerate with zero eigenvalues generated by the primitive $m$ roots of unity and according to \eqref{eq:GSD_gapless_formula} the contribution to the GSD is a mononomial in $N$ 
\ie\label{eq:rational_mod_0}
AN^{\sum_\alpha(\Gamma_\alpha-\Delta_\alpha)}~,
\fe
where $A$ is an $N$-indepedent constant and the sum in exponent is over roots that are primitive $m$th roots of unity. Combining \eqref{eq:rational_mod_1} and \eqref{eq:rational_mod_0}, we conclude that the pattern of the contribution to the GSD from the primitive $m$th roots of unity depends only on the value of $N$ modulo $m$.

If an iCS theory has only roots at the primitive $m$th roots of unity, then the determinant polynomial must be a power of the $m$th \emph{cyclotomic polynomial} $\Phi_m(u)$ with an integer coefficient $C$ (see Proposition \ref{prop:cyclotomic_min_poly} in Appendix \ref{app:cyclotomic} for more details)
\ie
D(u)=C u^{-\xi}\Phi_m(u)^{\Gamma}~.
\fe
The $m$th cyclotomic polynomial is defined as
\ie
\Phi_m(u)=\prod_{\substack{1\leq k\leq m\\ \text{gcd}(k,m)=1}}(u-e^{2\pi i k/m})~.
\fe
It is in fact an integer polynomial although not manifest in the definition.
The GSD of such an iCS theory is 
\ie
\text{GSD}=\begin{dcases}
A|C|^N N^{\gamma}~,\ \ &m\mid N
\\
\\
|C|^N\prod_{\substack{1\leq k\leq m\\\gcd(k,m)=1}}\left|2\sin\left(\frac{ k\pi N}{m}\right)\right|^{\Gamma}~,\ \ &m\nmid N 
\end{dcases}~,
\fe
where $\gamma=\sum_\alpha(\Gamma-\Delta_\alpha)$.
Restricting to a subsequence of $N$ with a fixed  $(N\text{ mod }m)\neq 0$, the GSD 
\begin{itemize}[leftmargin=*]
    \item 
    is a constant when $|C|=1$;
    \item
    grows exactly exponentially when $|C|\neq 1$.
\end{itemize}
On the other hand, on the subsequence of $N$ that are divisible by $m$, the GSD 
\begin{itemize}[leftmargin=*]
    \item 
    is a constant when $|C|=1$, $\gamma= 0$;
    \item
    grows polynomially when $|C|= 1$, $\gamma\neq 0$;
    \item 
    grows exponentially when $|C|\neq1$, $\gamma=0$;
    \item 
    grows exponentially with a polynomial overhead when $|C|\neq 1$, $\gamma\neq 0$.
\end{itemize}

As an example, consider a theory with a tridiagonal $K$ matrix \eqref{eq:tridiagonal} with $(M_0,M_1)=(-1,1)$. The theory has roots at $u=\exp(\pm 2\pi i/6)$ with $\Gamma=\Delta=1$. Its GSD is plotted in Figure \ref{fig:-11} and is given by
\ie
\text{GSD}=\begin{cases}
1~,\ \ &6\mid N
\\
4\sin^2(\pi N/6)~,\ \ &6\nmid N
\end{cases}~,
\fe
which oscillates between $1,3,4,3,1,1$ periodically.

For a more sophisticated example, consider another theory with a tridiagonal $K$ matrix \eqref{eq:tridiagonal} with $(M_0,M_1)=(2,1)$. The theory has roots at $u=-1$ with $\Gamma=2$ and $\Delta=1$. Its GSD is plotted in Figure \ref{fig:21} and is given by
\ie
\text{GSD}=\begin{cases}
N~,\ \ &2\mid N
\\
4~,\ \ &2\nmid N
\end{cases}~,
\fe
which grows linearly for even $N$ but retains a constant for odd $N$.

\section{Condition for Foliation}\label{sec:foliation}

Foliated fracton orders \cite{Shirley:2017suz} provide a useful organizing principle for gapped fracton phases. Examples of foliated fracton orders include the X cube model \cite{Vijay:2016phm}, the checkerboard model \cite{Vijay:2016phm,Shirley:2018hkm}, the twisted foliated fracton order \cite{Shirley_2020}, etc. As defined in Definition \ref{def:foliated}, a foliated fracton order uses decoupled layers of two-dimensional gapped topological orders as resources to increase its system size. As a result, the GSD of a foliated fracton order grows exponentially with the linear system size $N$
\ie\label{eq:foliated_GSD}
\text{GSD}=A M^N~,
\fe
where $A$ is an $N$-independent constant and $M\in\mathbb{Z}$ is the GSD of the resource topological order. 

The simplest foliated iCS theory is the one with a diagonal $K$ matrix 
\ie\label{eq:diagonal}
K=\left(
\begin{array}{ccccc}
M
\\
&M\\
&&\ddots\\
&&&M
\end{array}
\right)~.
\fe 
It is a foliated fracton order as increasing the system size amounts to inserting a decoupled layer of $U(1)_M$ Chern-Simons theory. As a result, its $\text{GSD}=M^N$ grows exponentially with $N$. 

A more non-trivial example of foliated iCS theory was discussed in \cite{Shirley_2020}, which has a $K$ matrix of period two
\ie\label{eq:period2matrix}
K=
\begin{pmatrix}
    \ddots&&&\\
&0&2&-1&\\
&2&0&&\\
&-1 &&0&2&-1\\
&&&2&0&\\
&&&-1&&0&2 & -1\\
&&&&&2&0
\\
&&&&&-1 && 0\\
&&&&&&&&\ddots
\end{pmatrix}~.
\fe
The theory has an exponentially growing  $\text{GSD}=4^N$.
One can decouple the resource topological order from the iCS theory by performing a $\text{GL}(2N,\mathbb{Z})$ transformation $K\mapsto W^T KW$ with the $\text{GL}(2N,\mathbb{Z})$ matrix
\ie
W=
\setcounter{MaxMatrixCols}{12}
\begin{pmatrix}
    \ddots &  &  &  &  &  &  &  &  & &  \\
& 1 &  &  &  &  &  &  &  & &  \\
&  & 1 &  &  &  & 1 &  &  & &  \\
&  &  & 1 &  &  &  & -1 &  & &  \\
&  &  &  & 1 &  &  &  &  & &  \\
& -1 &  &  &  & 1 &  &  &  & &  \\
&  &  &  &  &  & 1 &  &  & &  \\
&  &  &  &  &  &  & 1 &  & &  \\
& -1 &  &  & 1 &  &  &  & 1 & &  \\
&  &  &  &  &  &  &  &  & \ddots 
\end{pmatrix}~.
\fe
$W$ is the identity matrix outside the region displayed above. This transformation yields a new $K$ matrix
\ie
W^TKW=
\begin{pmatrix}
    \ddots &  &  &  &  &  &  &  &  &  \\
& 0 & 2 &  &  &  &  & -1 &  &  \\
& 2 & 0 &  &  &  &  &  &  &  \\
&  &  & 0 & 2 & -1 & 0 &  &  &  \\
&  &  & 2 & 0 & 0 & 0 &  &  &  \\
&  &  & -1 & 0 & 0 & 2 &  &  &  \\
&  &  & 0 & 0 & 2 & 0 &  &  &  \\
& -1 &  &  &  &  &  & 0 & 2 &  \\
&  &  &  &  &  &  & 2 & 0 &  \\
&  &  &  &  &  &  &  &  & \ddots
\end{pmatrix}~.
\fe
There is a $4\times 4$ block decoupled from the rest of the matrix which takes the same form as the original $K$ matrix. The decoupled block is the resource layer, in this case a $\mathbb{Z}_2\times \mathbb{Z}_2$ twisted gauge theory. 

Despite the examples above, as we will show below, generically a gapped iCS theory is expected to be non-foliated, whose GSD does not take the form as \eqref{eq:foliated_GSD}. 

An interesting and important question is to determine which gapped iCS theories are foliated and which ones are not. 
Here, we propose a necessary condition for a gapped iCS theory to be foliated:
\begin{proposition}\label{prop:foliation}
    A gapped iCS theory is foliated only if its determinant polynomial $D(u)$ is a constant.
\end{proposition}
\noindent 
This proposition can be proven by comparing the GSD of the gapped iCS theory \eqref{eq:det} with the GSD of a foliated fracton order \eqref{eq:foliated_GSD}. If $D(u)$ were not a constant, then it has some roots $u_\alpha$ which renders the GSD formula \eqref{eq:det} a sum over exponentials of $N$. The sum includes at least two terms: an exponential with the largest base $|C|\prod_{|u_\alpha|>1}|u_\alpha|^{\Gamma_\alpha}$ and an exponential with the smallest base $|C|\prod_{|u_\alpha|<1}|u_\alpha|^{\Gamma_\alpha}$. Such a sum can never be expressed as a single exponential of $N$. Hence, the theory cannot be foliated.

This condition aligns with the expectations from braiding statistics. In a foliated iCS theory, charges are expected to have non-trivial braiding only when their separations in the $x_3$ direction are finite. This ensures that charges with non-trivial braiding can be grouped into anyons of the resource topological order, allowing them to be decoupled from the rest of the iCS theory using finite-depth local unitary circuit.  This is incompatible with the existence of non-unit roots of the determinant polynomial. It is because non-unit roots lead to long-ranged interactions in the braiding phase that decay exponentially with the separation in the $x_3$ direction.
Thus, a foliated iCS theory is expected to have a constant determinant polynomial.

We now check the condition in Proposition \ref{prop:foliation} against the two examples of foliated iCS theories. The diagonal $K$ matrix \eqref{eq:diagonal} has a constant $D(u)=M$, which is consistent with the condition.
The period two $K$ matrix \eqref{eq:period2matrix} has
\ie
P(u)=
\left(\begin{array}{cc}
    u+u^{-1} & 2 \\
    2 & 0
\end{array}\right)~,\quad D(u)=4~.
\fe
The constant $D(u)$ is again consistent with the condition.

Although we are uncertain whether Proposition \ref{prop:foliation} is sufficient, it can already be used to exclude foliated fracton orders in a large class of gapped iCS theories. For example, any gapped iCS theory with a $K$ matrix of period one cannot be a foliated fracton order unless its $K$ matrix is diagonal. This is because the determinant polynomial $D(u)=\sum_{k}M_{k}u^k$ of a non-diagonal $K$ matrix is necessarily non-constant. 

To test how close Proposition \ref{prop:foliation} is close to be sufficient, let us construct a class of $K$ matrices with constant $D(u)$. These $K$ matrices have period 2 and are generalizations of \eqref{eq:period2matrix}. They take the form
\ie\label{eq:Kmatrix_test}
K=
\begin{pmatrix}
    \ddots&&&\\
&0&m&k&\\
&m&0&&\\
&k &0&m&&k\\
&&&m&0&\\
&&&k&&0&m & k\\
&&&&&m&0
\\
&&&&&k && 0\\
&&&&&&&&\ddots
\end{pmatrix}~.
\fe
This matrix has
\ie
P(u)=\begin{pmatrix}
    ku+ku^{-1} & m
    \\
    m &0
\end{pmatrix}~,\quad D(u)=m^2~.
\fe
The iCS theories with these $K$ matrices are all foliated. Their Lagrangian are given by
\ie
\mathcal{L}=\sum_i\left(\frac{m}{2\pi} a_i \wedge db_i+\frac{k}{2\pi} a_i\wedge da_{i+1}\right)
\fe
The $\text{GL}(2N,\mathbb{Z})$ transformation that decouples a resource topological order is given by
\begin{alignat}{2}
&a_{2}\rightarrow a_{2}-a_{4}+a_{6}~,
\ \ \quad &&a_{3}\rightarrow a_{3}-a_{1}~,\nonumber
\\
&a_{4}\rightarrow a_{4}-a_{6}~,
\ \ \quad &&a_{5}\rightarrow a_{5}+a_{1}~,\nonumber
\\
&b_{1}\rightarrow b_{1}+b_{3}-b_{5}~,\ \ \quad &&b_{4}\rightarrow b_{4}+b_{2}~,\nonumber
\\
&b_{6}\rightarrow b_{6}+b_{4}~.
\end{alignat}
After the transformation, the Lagrangian becomes
\ie
\mathcal{L}&=\mathcal{L}_{\text{stack}}+\mathcal{L}_{\text{resource}}~,
\fe
where $\mathcal{L}_{\text{stack}}$ describes the same iCS theory with the system size decreases by 4 layers,
\ie
\mathcal{L}_{\text{stack}}&=\sum_{i\leq1}\left(\frac{m}{2\pi} a_i \wedge db_i +\frac{k}{2\pi} a_{i-1}\wedge da_{i}\right)+\frac{k}{2\pi} a_1\wedge da_6
\\
&+\sum_{i\geq6}\left(\frac{m}{2\pi} a_i \wedge db_i+\frac{k}{2\pi} a_i\wedge da_{i+1}\right)~,
\fe
while $\mathcal{L}_{\text{resource}}$ describes a decoupled topological order
\ie
\mathcal{L}_{\text{resource}}=\sum_{i=2}^5\frac{m}{2\pi} a_i \wedge db_i+\frac{k}{2\pi} (a_2\wedge a_3+a_4\wedge a_5)~.
\fe
This means that the iCS theory \eqref{eq:Kmatrix_test} is indeed foliated.

\begin{acknowledgements}
We are grateful to Daren Chen, Meng Cheng, Roman Geiko, Yi Ni and Yiyue Zhu for discussions. X.M. and X.C. are supported by the National Science Foundation under award number DMR-1654340, the Simons collaboration on ``Ultra-Quantum Matter'' (grant number 651440), the Simons Investigator Award (award ID 828078) and the Institute for Quantum Information and Matter at Caltech. X.C. is also supported by the Walter Burke Institute for Theoretical Physics at Caltech.  H.T.L. is supported in part by a Croucher fellowship from the Croucher Foundation, the Packard
Foundation and the Center for Theoretical Physics at MIT. The authors of this paper were ordered
alphabetically.
\end{acknowledgements}

\appendix

\section{Supplementary mathematics}
\label{app:math}

In this appendix, we discuss the mathematics necessary for proving the two propositions at the end of Section~\ref{sec:poly}. Most of the statements can be found in standard mathematics textbooks such as \cite{cohn2000classic, Hecke1981}. Appendix \ref{app:1},~\ref{app:2},~\ref{app:3} discusses the background materials, Appendix \ref{app:cyclotomic} proves Proposition \ref{prop:primitive_roots} using the properties of cyclotomic polynomials, and Appendix \ref{app:prop2} proves Proposition \ref{prop:all_rational}.

\subsection{Definitions}\label{app:1}

Let $\mathbb{Z}[u]$ (resp.\ $\mathbb{Q}[u]$) be the ring of polynomials in $u$ with integer (resp.\ rational) coefficients. Although in the main text we have used \textit{Laurent} polynomials, none of the statements in this appendix depends on whether terms with negative powers are allowed. Therefore, we will only consider ordinary polynomials here. Take
\ie
\alpha(u)=a_n u^n+\cdots+a_0\in\mathbb{Z}[u]~,
\fe
where $a_n\neq0$. We say that $\alpha(u)$ is \textit{monic} if $a_n=1$. The \textit{content} of $\alpha(u)$ is
\ie
c(\alpha)=\text{gcd}\{a_k\}~.
\fe
We say that $\alpha(u)$ is \textit{primitive} if $c(\alpha)=1$.\footnote{Note that this is different from ``primitive $m$th root of unity''.} We say that $\alpha(u)$ is \textit{reducible} if it can be written as
\ie
\alpha(u)=\beta(u)\gamma(u)~,
\fe
where neither of $\beta(u)$ and $\gamma(u)$ is invertible in $\mathbb{Z}[u]$. Otherwise, $\alpha(u)$ is \textit{irreducible}. In $\mathbb{Z}[u]$, the only invertible elements are $\pm1$. We have the following lemma:
\begin{lemma}\label{lemma:monic_division}
    Let $\alpha(u),\beta(u)\in\mathbb{Z}[u]$ with $\alpha(u)$ monic. If
    \ie
    \gamma(u)=\frac{\beta(u)}{\alpha(u)}\in\mathbb{Q}[u]~,
    \fe
    then actually $\gamma(u)\in\mathbb{Z}[u]$.
\end{lemma}
\noindent This lemma is clear from the procedure of long division. We will use the lemma several times in this appendix. Due to its simplicity, we will not always quote the lemma explicitly when it is used.

Content and reducibility can also be defined in $\mathbb{Q}[u]$, where the convention for gcd is e.g.
\ie
\text{gcd}\left\{\frac{1}{2},\frac{1}{3}\right\}=\frac{1}{6}~,
\fe
and the invertible elements are $q\in\mathbb{Q}$, $q\neq0$. We have the following lemmas:
\begin{lemma}\label{lemma:Gauss_1}
    If $\alpha(u),\beta(u)\in\mathbb{Q}[u]$, then the content of the product $\alpha(u)\beta(u)$ is $c(\alpha\beta)=c(\alpha)c(\beta)$. In particular, the product of two primitive polynomials is also primitive.
\end{lemma}
\begin{lemma}\label{lemma:Gauss_2}
    If $\alpha(u)\in\mathbb{Z}[u]$ is primitive and irreducible in $\mathbb{Z}[u]$, then it is also irreducible when viewed as an element of $\mathbb{Q}[u]$.
\end{lemma}
\noindent These lemmas are known collectively as Gauss's lemma.

\subsection{Algebraic Number and Minimal Polynomial}\label{app:2}

Now take a number $z\in\mathbb{C}$. We say that $z$ is an \textit{algebraic number} if it satisfies some polynomial $\alpha(u)\in\mathbb{Q}[u]$, i.e.\ $\alpha(z)=0$. An equivalent definition is to replace $\mathbb{Q}[u]$ by $\mathbb{Z}[u]$, since if $\alpha(z)=0$ then $b\alpha(z)=0$ for any integer $b\neq0$. Define the \textit{annihilator} of an algebraic number $z$ as the set
\ie
\text{Ann}(z)=\left\{\alpha(u)\in \mathbb{Q}[u]\,|\, \alpha(z)=0\right\}~,
\fe
and the \textit{minimal polynomial} $\alpha_z(u)$ of $z$ as the element of $\text{Ann}(z)$ of the minimal degree. For example, if $z=1/2$ then we can take $\alpha_z(u)=2u-1$. Note that the minimal polynomial is always defined up to multiplication by $q\in\mathbb{Q}$, $q\neq0$. Aside from this ambiguity, $\alpha_z(u)$ is unique, which follows from the fact that
\ie\label{eq:min_poly_gcd}
\alpha_z(u)=\text{gcd}\left(\text{Ann}(z)\right)~,
\fe
where we are taking the gcd of polynomials in $\mathbb{Q}[u]$. To give an example, $\text{gcd}\{\beta(u),\gamma(u)\}$ is the polynomial of the maximal degree that divides both $\beta(u)$ and $\gamma(u)$. We can prove \eqref{eq:min_poly_gcd} by contradiction: Suppose that $\alpha_z(u)$ does not divide some $\beta(u)\in\text{Ann}(z)$. A property of the gcd is that there exist $\phi(u),\chi(u)\in\mathbb{Q}[u]$ such that
\ie
\gamma(u)&=\text{gcd}\{\alpha_z(u),\beta(u)\}\\
&=\phi(u)\alpha_z(u)+\chi(u)\beta(u)~.
\fe
This then implies that $\gamma(u)\in\text{Ann}(z)$. Since $\alpha_z(u)$ does not divide $\beta(u)$, their gcd $\gamma(u)$ has a smaller degree than $\alpha_z(u)$, contradicting the definition of $\alpha_z(u)$. Similarly, $\alpha_z(u)$ is irreducible, since any non-trivial factor of $\alpha_z(u)$ would have a lower degree than $\alpha_z(u)$. Theses facts imply the following corollary:
\begin{corollary}\label{ref:lemma_min_poly}
    The minimal polynomial of an algebraic number $z$ is (up to multiplication by $q\in \mathbb{Q}$, $q\neq0$) the unique irreducible element of $\text{Ann}(z)$.
\end{corollary}
\noindent This lemma will be used later to prove Proposition~\ref{prop:cyclotomic_min_poly} and hence Proposition~\ref{prop:primitive_roots}.

\subsection{Algebraic Integer}\label{app:3}

We say that $z\in\mathbb{C}$ is an \textit{algebraic integer} if it satisfies some monic polynomial $\alpha(u)\in\mathbb{Z}[u]$. Note that we are restricting the polynomials from $\mathbb{Q}[u]$ to $\mathbb{Z}[u]$ here. An algebraic integer is always an algebraic number, but not vice versa. We have the following lemma:
\begin{lemma}
    A number $z\in\mathbb{C}$ is an algebraic integer if and only if its minimal polynomial in $\mathbb{Z}[u]$ is monic.
\end{lemma}
\noindent In $\mathbb{Z}[u]$, the minimal polynomial is also required to be primitive besides having the minimal degree. The ``if'' part of the lemma is obvious. For the ``only if'' part, take a monic polynomial $\beta(u)\in\text{Ann}(z)$ and let
\ie
\gamma(u)=\frac{\beta(u)}{\alpha_z(u)}~.
\fe
By Lemma~\ref{lemma:Gauss_1},
\ie
c(\gamma)=\frac{c(\beta)}{c(\alpha_z)}=1~,
\fe
so $\gamma(u)\in\mathbb{Z}[u]$. When multiplying two polynomials $\alpha_z(u)$ and $\gamma(u)$, the leading coefficient of the product is the product of the leading coefficients. Since $\beta(u)$ is monic by assumption, so are $\alpha_z(u)$ and $\gamma(u)$. We have the following corollary:
\begin{corollary}\label{cor:rational_integer}
    A rational number $q\in\mathbb{Q}$ is an algebraic integer if and only if it is an ordinary integer.
\end{corollary}
\noindent To see this, write $q=r/s$ where $r,s\in\mathbb{Z}$ are coprime. Then the minimal polynomial of $q$ is
\ie
\alpha_q(u)=su-r~,
\fe
which is monic if and only if $s=1$. This corollary will also be used to prove Proposition~\ref{prop:cyclotomic_min_poly}.

Furthermore, it can be shown that algebraic integers form a ring.

\subsection{Cyclotomic Polynomial and Proof of Proposition~\ref{prop:primitive_roots}}\label{app:cyclotomic}
Having prepared the basics, we now begin a discussion of roots of unity which eventually leads to Propositions~\ref{prop:primitive_roots} and \ref{prop:all_rational}. Recall that an $m$th root of unity $e^{2\pi i k/m}$ is called \textit{primitive} if $k$ is coprime with $m$.\footnote{Note that this is difference from ``primitive polynomial''.} We write
\ie\label{eq:cyclotomic_def}
C_m=\{k\in\mathbb{Z}_m\,|\, \text{gcd}(k,m)=1\}~,
\fe
and define the $m$th \textit{cyclotomic polynomial} $\Phi_m(u)$ as
\ie
\Phi_m(u)=\prod_{k\in C_m}(u-\omega^k)~,
\fe
where $\omega=e^{2\pi i/m}$. Equivalently,
\ie\label{eq:cyclotomic_alt}
\Phi_m(u)=\dfrac{u^m-1}{\prod\limits_{k\notin C_m}(u-\omega^k)}~.
\fe
The denominator of \eqref{eq:cyclotomic_alt} is a product of cyclotomic polynomials $\Phi_{m'}(u)$ where $m'<m$. Thus by induction on $m$ and Lemma~\ref{lemma:monic_division}, it is straightforward to show that $\Phi_m(u)\in\mathbb{Z}[u]$. Clearly, $\Phi_m(u)$ is monic and is satisfied by all primitive $m$th roots of unity. In fact:
\begin{proposition}\label{prop:cyclotomic_min_poly}
    All primitive $m$th roots of unity share the same minimal polynomial, which is $\Phi_m(u)$.
\end{proposition}
\noindent By Corollary~\ref{ref:lemma_min_poly}, it is enough to show that $\Phi_m(u)$ is irreducible. We give a proof found in \cite{Weintraub2013}. Suppose that $\beta(u)\in\mathbb{Z}[u]$ divides $u^m-1$. Let $u_0$ be a root of $\beta(u)$ (and hence an $m$th root of unity), and $p$ a prime that does not divide $m$. We claim that $u_0^p$ is also a root of $\beta(u)$. If this claim is true, then we can choose $u_0$ to be a primitive $m$th root of unity and $\beta(u)$ its minimal polynomial $\alpha_{u_0}(u)$. By basic number theory, every primitive $m$th root of unity is of the form $u_0^s$ for some $s\in\mathbb{Z}$ coprime with $m$. We can then apply the claim several times using the factorization $s=p_1^{k_1}\cdots p_n^{k_n}$, where none of the prime factors $p_i$ divides $m$. Therefore, $\alpha_{u_0}(u)$ has all primitive $m$th roots of unity as its roots, so $\Phi_m(u)$ divides $\alpha_{u_0}(u)$. However, the minimal polynomial $\alpha_{u_0}(u)$ of $u_0$ is irreducible, so actually $\alpha_{u_0}(u)=\Phi_m(u)$. Since this argument works for all primitive $m$th roots of unity, Proposition~\ref{prop:cyclotomic_min_poly} is proved.

To prove the above claim, we start by defining an auxiliary quantity
\ie
\Delta&=\prod_{j<k}\left(\omega^j-\omega^k\right)^2=\pm\prod_{j\neq k}\left(\omega^j-\omega^k\right)\\
&=\pm\prod_j\left[\omega^j\prod_{l=1}^{m-1}\left(1-\omega^l\right)\right]=\pm\prod_j\left(m\omega^j\right)=\pm m^m~,
\fe
where all labels take values in $\mathbb{Z}_m$. When going from the fourth to the fifth line, we set $x=1$ in the polynomial
\ie
f(x)=\prod_{l=1}^{m-1}\left(x-\omega^l\right)=\frac{x^m-1}{x-1}=\sum_{l=0}^{m-1}x^l~,
\fe
which gives $f(1)=m$. The quantity $\Delta$ is known as the \textit{discriminant} of the polynomial $u^m-1$, but we do not need a formal definition of this concept here. Suppose that the claim is false, i.e.\ that $u_0^p$ is not a root of $\beta(u)$. Without loss of generality, $\beta(u)$ is monic. We can then use the roots $u_i$ of $\beta(u)$ to write
\ie
\beta(u_0^p)=\prod_i (u_0^p-u_i)\neq0~.
\fe
Thus $\beta(u_0^p)$ is a product of differences of distinct $m$th roots of unity. Furthermore, $\beta(u_0^p)$ is also an algebraic integer since algebraic integers form a ring. Therefore, $\beta(u_0^p)$ divides $\Delta$ as algebraic integers, i.e.\ $\Delta/\beta(u_0^p)$ is an algebraic integer. Now we can apply the multinomial theorem to $\beta(u)^p$ to show that
\ie
\beta(u_0^p)=\beta(u_0)^p=0\ (\text{mod }p)~.
\fe
Here, ``equality mod $p$'' means that the difference of the two sides is $p$ times an algebraic integer. Thus $p$ divides $\beta(u_0^p)$ as algebraic integers. To explain this step in more detail, suppose that $\beta(u)=\sum_kb_ku^k$, where $b_k$ are ordinary integers. In the multinomial expansion of $\beta(u)^p$, every cross term has a multinomial coefficient that is divisible by $p$, so
\ie
\beta(u)^p=\sum_k b_k^p u^{kp}\ (\text{mod }p)~.
\fe
By applying Fermat's little theorem to the ordinary integer $b_k$, we have $b_k^p=b_k$ (mod $p$). Therefore,
\ie
\beta(u)^p=\beta(u^p)\ (\text{mod }p)~.
\fe
Consequently, $p$ divides $\Delta$ as algebraic integers, i.e.
\ie
\frac{\Delta}{p}=\frac{\Delta}{\beta(u_0^p)}\frac{\beta(u_0^p)}{p}
\fe
is an algebraic integer. However, Corollary~\ref{cor:rational_integer} says that the rational number $\Delta/p$ is an algebraic integer if and only if it is an ordinary integer. The conclusion is that $p$ divides $\Delta=\pm m^m$ as ordinary integers, which contradicts the assumption that $p$ does not divide $m$. This proves the claim and hence Proposition~\ref{prop:cyclotomic_min_poly}.

Proposition~\ref{prop:primitive_roots} can be derived easily from Proposition~\ref{prop:cyclotomic_min_poly}. Suppose that a primitive $m$th root of unity $u_i$ is a root of $D(u)$ with multiplicity $\Gamma_i$. Since $D(u)\in\text{Ann}(u_i)$, the cyclotomic polynomial $\Phi_m(u)$ must divide $D(u)$ (up to some trivial technicality we can pretend $D(u)\in\mathbb{Z}[u]$). If $\Gamma_i>1$, then the same argument can be applied to $D(u)/\Phi_m(u)$. By induction, $\Phi_m(u)^{\Gamma_i}$ divides $D(u)$. Therefore, all primitive $m$th roots of unity are roots of $D(u)$ with the same multiplicity $\Gamma_i$.

\subsection{Proof of Proposition \ref{prop:all_rational}}\label{app:prop2}

Finally, we prove Proposition~\ref{prop:all_rational} following \cite{Kronecker1857}. Let $u_1,\ldots,u_n$ be the roots of $D(u)$. Here, a repeated root of multiplicity $\Gamma$ is listed $\Gamma$ times as different roots. For $k=1,\ldots,n$, let $e_k(x_1,\ldots,x_n)$ be the $k$th elementary symmetric polynomial in $n$ variables. For example,
\ie
e_1(x_1,\ldots,x_n)&=\sum_i x_i~,\\
e_2(x_1,\ldots,x_n)&=\sum_{i<j} x_ix_j~,\\
e_3(x_1,\ldots,x_n)&=\sum_{i<j<k} x_ix_jx_k~.
\fe
By Vieta's theorem, $e_k(u_1,\ldots,u_n)\in\mathbb{Z}$ for all $k$. Now for each integer $s>0$, define a polynomial
\ie
\beta_s(u)=\prod_{k=1}^n(u-u_k^s) =\sum_{k=0}^nb_{sk}u^k~.
\fe
Each coefficient $b_{sk}$ is a symmetric polynomial in the $n$ variables $u_1,\ldots,u_n$ with integer coefficients. By the fundamental theorem of symmetric polynomials, $b_{sk}$ can be written uniquely as
\ie
b_{sk}=\gamma_{sk}(e_1(u_1,\ldots,u_n),\ldots,e_n(u_1,\ldots,u_n))~,
\fe
where $\gamma_{sk}(x_1,\ldots,x_n)$ is a polynomial in $n$ variables with integer coefficients. The conclusion is that $b_{sk}\in\mathbb{Z}$. Also, since $|u_k|=1$, we have the bound
\ie
|b_{sk}|\le\binom{n}{k}~.
\fe
As a result, the list $(b_{s1},\ldots,b_{sn})$ can only take finitely many values, so there exist integers $0\le g<h$ such that $\beta_{2^g}(u)=\beta_{2^h}(u)$. Therefore, the list $(u_1^{2^h},\ldots,u_n^{2^h})$ is a permutation of the list $(u_1^{2^g},\ldots,u_n^{2^g})$. Let $c=2^g$ and $d=2^{h-g}$. Then this permutation $\sigma\in S_n$ of the list $(u_1^c,\ldots,u_n^c)$ is achieved by raising each element to the power $d$. Let $r$ be the order of $\sigma$ in the group $S_n$. Since the action of raising to the power $d$ repeated $r$ times is trivial, we see that $u_k^{rcd}=1$ for all $k$. This shows that all roots of $D(u)$ are rational roots.

\section{Fusion Group of iCS Theories with Period One}\label{app:one-form sym}

In this appendix, we study the fusion group or relatedly the one-form symmetry group of iCS theories with period $L=1$. Consider a period one $K$ matrix with
\ie
D(u)=\sum_{k=-\xi}^\xi M_ku^k~,
\fe
where $M_\xi\neq0$. We will prove the following bound on the fusion group:
\begin{proposition}\label{prop:fusion}
    If the $K$ matrix has period one and $\gcd\{M_k\}=1$, then the fusion group $G$ has at most $2\xi$ cyclic components.
\end{proposition}
\noindent 
As defined in Appendix~\ref{app:1}, $\gcd\{M_k\}=1$ means that $D(u)$ is a primitive polynomial. In the proposition, we always merge the cyclic components in the fusion group $G$ as much as possible. For example, $\mathbb{Z}_6=\mathbb{Z}_2\times\mathbb{Z}_3$ is viewed as having one cyclic component instead of two.  
To demonstrate the proposition, consider a period one $K$ matrix with $D(u)=u+3+u^{-1}$. The fusion group is $G=\mathbb{Z}_{F_N}\times\mathbb{Z}_{5F_N}$ which has two cyclic components, where $F_N$ is the $N$th Fibonacci number \cite{Ma:2020svo}.

As explained in Section~\ref{sec:GSD}, the fusion group is
\ie
G=\prod_{i=1}^{N}\mathbb{Z}_{r_i}~,
\fe
where $r_i$ are the diagonal entries of the Smith normal form \eqref{eq:SNF} of $K$, and we use the convention that $\mathbb{Z}_0$ is the trivial group. Therefore, it suffices to show that
\ie
r_1=\cdots=r_{N-2\xi}=1~.
\fe
Since $r_i$ are integers, this is equivalent to showing that
\ie\label{eq:fusion_group_product}
\prod_{i=1}^{N-2\xi}r_i=1~.
\fe
This product is suitable for the current problem due to the following fact:
\ie\label{eq:minor_theorem}
\text{gcd}\left\{\text{minor of } K \text{ of order $m$}\right\}=\prod_{i=1}^{m}r_i~,
\fe
where a \textit{minor} of $K$ of order $m$ is the determinant of an $m\times m$ submatrix of $K$. By \eqref{eq:minor_theorem}, the problem becomes showing that
\ie\label{eq:minor_fusion}
\text{gcd}\left\{\text{minor of } K \text{ of order $N-2\xi$}\right\}=1~.
\fe
Therefore, we will pick some submatrices of $K$ obtained by deleting $2\xi$ rows and $2\xi$ columns, and show that their determinants are coprime as a whole. 

First, delete the first $2\xi$ columns and the last $2\xi$ rows of the $K$ matrix and consider the submatrix
\ie
T_{\xi}=\begin{pmatrix}
M_\xi\\
M_{\xi-1} & M_\xi\\
\vdots & \ddots & \ddots\\
0 & \cdots & M_{\xi-1} & M_\xi
\end{pmatrix}~.
\fe
Clearly $\det(T_\xi)=M_\xi^{N-2\xi}$ since $T_\xi$ is lower-triangular. Next, instead of deleting the first $2\xi$ columns of the $K$ matrix, delete the first $2\xi-1$ columns and the last column and consider the submatrix
\ie
T_{\xi-1}=\begin{pmatrix}
M_{\xi-1} & M_\xi\\
M_{\xi-2} & M_{\xi-1} & M_\xi\\
\vdots & \ddots & \ddots & \ddots\\
0 & \cdots & M_{\xi-2} & M_{\xi-1} & M_\xi\\
0 & 0 & \cdots & M_{\xi-2} & M_{\xi-1}
\end{pmatrix}~.
\fe
When expanding $\det(T_{\xi-1})$ into a sum of terms using the definition of determinant, a term either is $M_{\xi-1}^{N-2\xi}$, or contains a factor of $M_\xi$. Therefore, 
\ie\label{eq:modulo}
\det(T_{\xi-1})=M_{\xi-1}^{N-2\xi}\ (\text{mod } M_\xi)~.
\fe
Suppose there exists a prime $p$ that divides both $\det(T_\xi)$ and $\det(T_{\xi-1})$. It then also divides $M_\xi$ since $\det(T_{\xi})=M_\xi^{N-2\xi}$. Using the relation \eqref{eq:modulo}, we can further deduce that $M_{\xi-1}^{N-2\xi}$ is divisible by $p$ and so does $M_{\xi-1}$.

Likewise, we can define submatrices $T_{\xi-2},\ldots,T_0$, and show inductively that if $p$ divides all $\det(T_i)$ it must also divide all $M_i$. This however contradicts our assumption that $\text{gcd}\{M_k\}=1$, so such a prime $p$ cannot exist and we must have $\text{gcd}\{\det(T_k)\}=1$. Since a subset of minors is already coprime, the set \eqref{eq:minor_fusion} of all minors of order $N-2\xi$ is also coprime, and hence Proposition~\ref{prop:fusion} is proved.

More generally, if $b=\text{gcd}\{M_k\}>1$, we can apply the proposition to the integer matrix $K/b$. After multiplying $b$ back, we see that the fusion group $G$ has at least $N-2\xi$ cyclic components that are $\mathbb{Z}_b$. In general, the other $2\xi$ cyclic components still grow (asymptotically) exponentially with $N$, affirming the non-foliated nature of the theory unless $K$ is diagonal.

\bibliographystyle{apsrev4-1}
\bibliography{ref}

\end{document}